\documentclass[conference]{IEEEtran}
\usepackage{amsmath,cite,bm,amsfonts,amssymb,graphicx,color,mathtools,setspace,comment,multirow,xfrac}
\usepackage{epstopdf}
\usepackage{graphicx}
\usepackage{caption}
\usepackage{subcaption}
\usepackage{color}

\newcommand{\bE}{\mathbf{E}}

\newcommand{\bh}{\mathbf{h}}
\newcommand{\bH}{\mathbf{H}}
\newcommand{\bI}{\mathbf{I}}

\newcommand{\bR}{\mathbf{R}}

\newcommand{\bW}{\mathbf{W}}
\newcommand{\bX}{\mathbf{X}}

\newcommand{\raisecaption}{\vspace{-0.7cm}}

\begin{document}

\title{On the Impact of Antenna Topologies for Massive MIMO Systems}
\author{\IEEEauthorblockA{Callum T. Neil\IEEEauthorrefmark{1},
											Mansoor Shafi\IEEEauthorrefmark{2},
											Peter J. Smith\IEEEauthorrefmark{3},
											Pawel A. Dmochowski\IEEEauthorrefmark{1}
											}
\IEEEauthorblockA{\IEEEauthorrefmark{1}
\small{School of Engineering and Computer Science, Victoria University of Wellington, Wellington, New Zealand}}
\IEEEauthorblockA{\IEEEauthorrefmark{2}
\small{Spark New Zealand, Wellington, New Zealand}}
\IEEEauthorblockA{\IEEEauthorrefmark{3}
\small{Department of Electrical and Computer Engineering, University of Canterbury, Christchurch, New Zealand}}
\IEEEauthorblockA{\small{email:\{pawel.dmochowski,callum.neil\}@ecs.vuw.ac.nz,~p.smith@elec.canterbury.ac.nz,~mansoor.shafi@spark.co.nz}}}

\maketitle

\begin{abstract}
Approximate expressions for the spatial correlation of cylindrical and uniform rectangular arrays (URA) are derived using measured distributions of angles of departure (AOD) for both the azimuth and zenith domains. We examine massive multiple-input-multiple-output (MIMO) convergence properties of the correlated channels by considering a number of convergence metrics. The per-user matched filter (MF) signal-to-interference-plus-noise ratio (SINR) performance and convergence rate, to respective limiting values, of the two antenna topologies is also explored.
\end{abstract}
\section{Introduction}
\label{sec:Introduction}
In order to meet demands for increased system capacity with limited spectral resources, there is a broad consensus that this can only be achieved via a large increase in system degrees of freedom (d.o.f.) \cite{HOYDIS2,LI,BJORNSON}. Consequently, massive multiple-input-multiple-output (MIMO) systems are being investigated \cite{MARZETTA,NGO,HOYDIS}, where the number of antennas are scaled up by at least an order of magnitude relative to current base station (BS) deployments. In the case of massive MIMO, fast-fading effectively averages out \cite{RUSEK,SMITH,GAO}, simplifying system analysis. Practical implementation of large numbers of antennas required for massive MIMO, typically in confined antenna array dimensions, however, results in reduced inter-element spacing adversely impacting system performance by way of spatial correlation \cite{SHIU}. Spatial correlation models for wireless systems are thus essential for accurate theoretical performance analysis and guarantees.
\par
An approximate spatial correlation model is presented in \cite{FORENZA} for clustered MIMO channels, deriving closed-form (CF) expressions for a uniform linear array (ULA) and a uniform circular array (UCA). This model however does not consider the zenith domain, necessary for accurate performance analysis. A correlation matrix is derived in \cite{YING} which considers both the azimuth and zenith domains, from which it is shown that the correlation matrix can be written as the Kronecker product of each domains correlation matrix. We propose to extend the works of \cite{YING} to massive MIMO antenna topologies: namely for uniform rectangular array (URA) and cylindrical antenna arrays, while incorporating Third Generation Partnership Project (3GPP) three-dimensional (3D) urban macro cell (UMa) environment measured distributions \cite{3GPP}. With the likelihood of smaller cell sizes and increased line-of-sight (LOS) propagation for next generation wireless systems, we make a small angle approximation to derive correlation expressions.
\par
The contributions of this paper are as follows:
\begin{itemize}
	\item We derive approximations for the spatial correlation of a URA and cylindrical antenna arrays using measured probability density functions (PDFs) of the AOD in both azimuth and zenith domains.
	\item We show, via simulation, the impact of antenna array topologies on the convergence rate of the channel to massive MIMO channel properties. This is shown via a number of channel convergence metrics.
	\item We show, via simulation, the impact of antenna array topologies on performance and rate of convergence, to respective limiting values, of matched filter (MF) per-user signal-to-interference-plus-noise ratio (SINR).
	\item We show that antenna arrays introduce a correlation structure which does not vary between topology. As a result, massive MIMO convergence metrics do not show any sensitivity to antenna topology, but do show sensitivity to the presence of correlation - which destroys the onset of massive MIMO properties.
\end{itemize}
%
\section{System Model}
\label{sec:SystemModel}
\subsection{System Description}
\label{subsec:System_Description}
We consider a multi-user (MU) massive MIMO downlink (DL) system with $M$ transmit antennas jointly serving a total of $K$ single-antenna users. We assume time division duplex (TDD) operation with uplink (UL) pilots enabling the transmitter to estimate the DL channel. The $M\times K$ channel matrix, $\bH$, is given by \cite{PAULRAJ}
\begin{equation}
	\bH = \bR_{\textrm{t}}^{\frac{1}{2}}\bH_{\textrm{iid}} \label{channel_matrix},
\end{equation}
where $\bH_{\textrm{iid}}$ is the $M\times K$ independent and identically distributed (i.i.d.) channel matrix with $\mathcal{CN}(0,1)$ entries, accounting for small-scale Rayleigh fading, and $\bR_{\textrm{t}}$ is the $M\times M$ spatial correlation matrix. We consider a cross-polarized (x-pol) antenna configuration, with the spatial correlation matrix modeled via \cite{PAULRAJ}
\begin{equation}
	\bR_{\textrm{t}} = \bX_{\textrm{pol}} \odot \bR,
\end{equation}
where the $M\times M$ matrix $\bR$ is the co-polarized (co-pol) spatial correlation matrix, $\odot $ represents the Hadamard product and $\bX_{\textrm{pol}}$ is the $M\times M$ x-pol matrix given by
\begin{equation}
	\bX_{\textrm{pol}} = \mathbf{1}_{M/2} \otimes \begin{bmatrix}
					1 & \sqrt{\delta } \\
					\sqrt{\delta } & 1 
				\end{bmatrix},
\end{equation}
where $\mathbf{1}_{M/2}$ is a $\frac{M}{2}\times \frac{M}{2}$ matrix of ones, $\delta $ denotes the cross-correlation between the two antenna elements in the x-pol configuration and $\otimes $ represents the Kronecker product.
\subsection{Channel Model}
\label{subsec:Channel_Model}
\begin{figure}
        \centering
        \begin{subfigure}[b]{0.22\textwidth}
        	\centering\includegraphics[trim=6cm 12cm 7cm 9.5cm,clip,width=1\columnwidth]{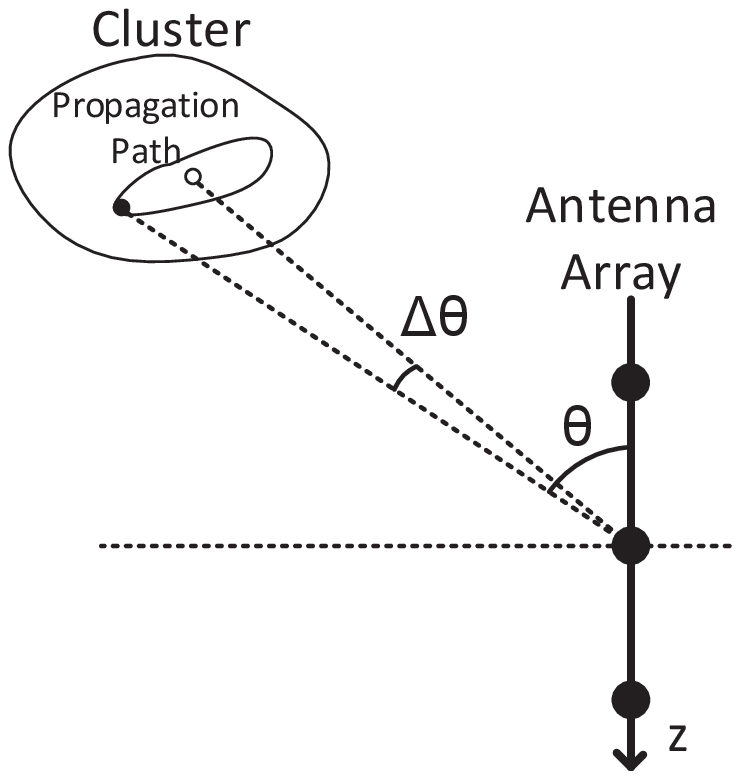}
                \caption{}
                \label{fig:ray_based_cluster_model}
        \end{subfigure}
        ~
        \begin{subfigure}[b]{0.22\textwidth}
		\centering\includegraphics[trim=5.5cm 10.5cm 4.5cm 7.5cm,clip,width=1\columnwidth]{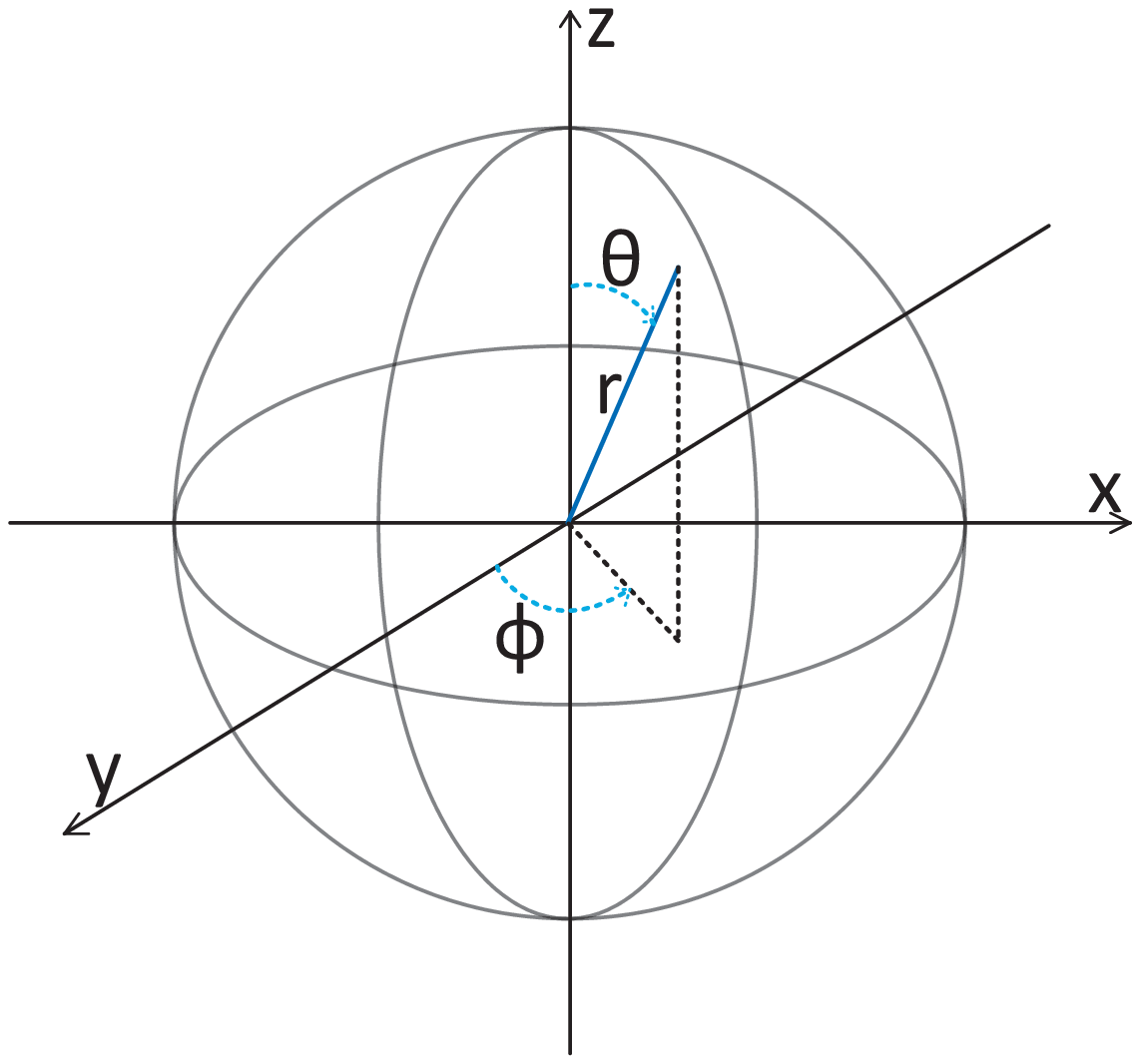}
		\caption{}
                \label{fig:channel_model_geometry}
        \end{subfigure}
	\caption{(a) Cluster Model. (b) Channel Model Geometry.}
\end{figure}

We consider a clustered channel model, shown in Figure \ref{fig:ray_based_cluster_model}, where we show the zenith AOD offset, $\Delta \theta $, of the wavefront relative to the mean zenith AOD of the cluster, $\theta $. In this scenario, the linear antenna array is located on the $z$ axis. Similarly, if the antenna array was positioned on the $y$ axis, Figure \ref{fig:ray_based_cluster_model} would describe the offset azimuth AOD, $\Delta \phi $, relative to the mean azimuth AOD of the cluster, $\phi $. The channel model geometry is clarified in Figure \ref{fig:channel_model_geometry}.
\par
\begin{figure}
        \centering
        \begin{subfigure}[b]{0.22\textwidth}
        	\centering\includegraphics[trim=7cm 14cm 6cm 6.67cm,clip,width=1\columnwidth]{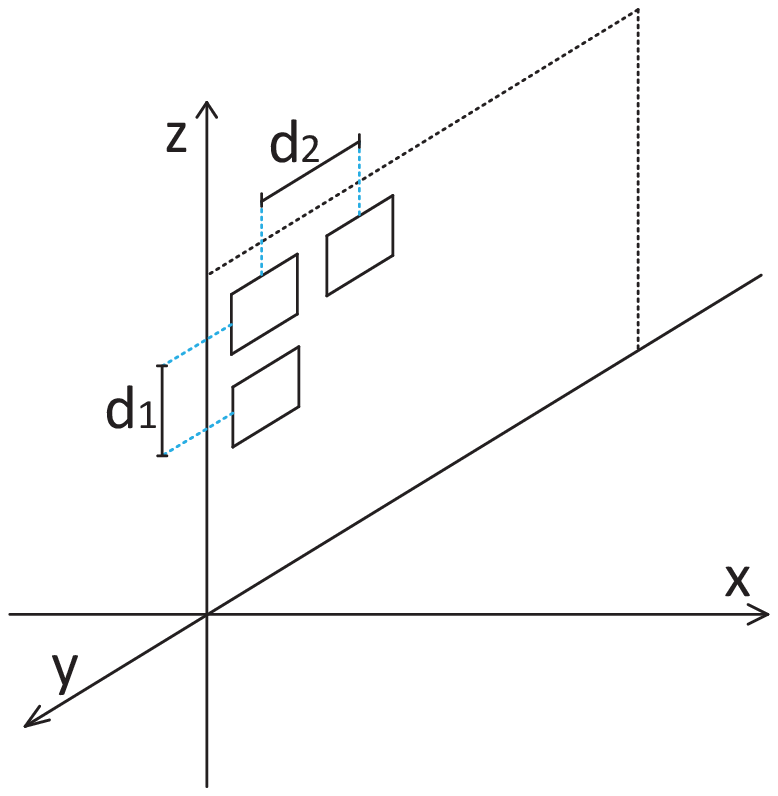}
                \caption{}
                \label{fig:ura}
        \end{subfigure}
        ~
        \begin{subfigure}[b]{0.22\textwidth}
		\centering\includegraphics[trim=5cm 5cm 4cm 7.3cm,clip,width=0.9\columnwidth]{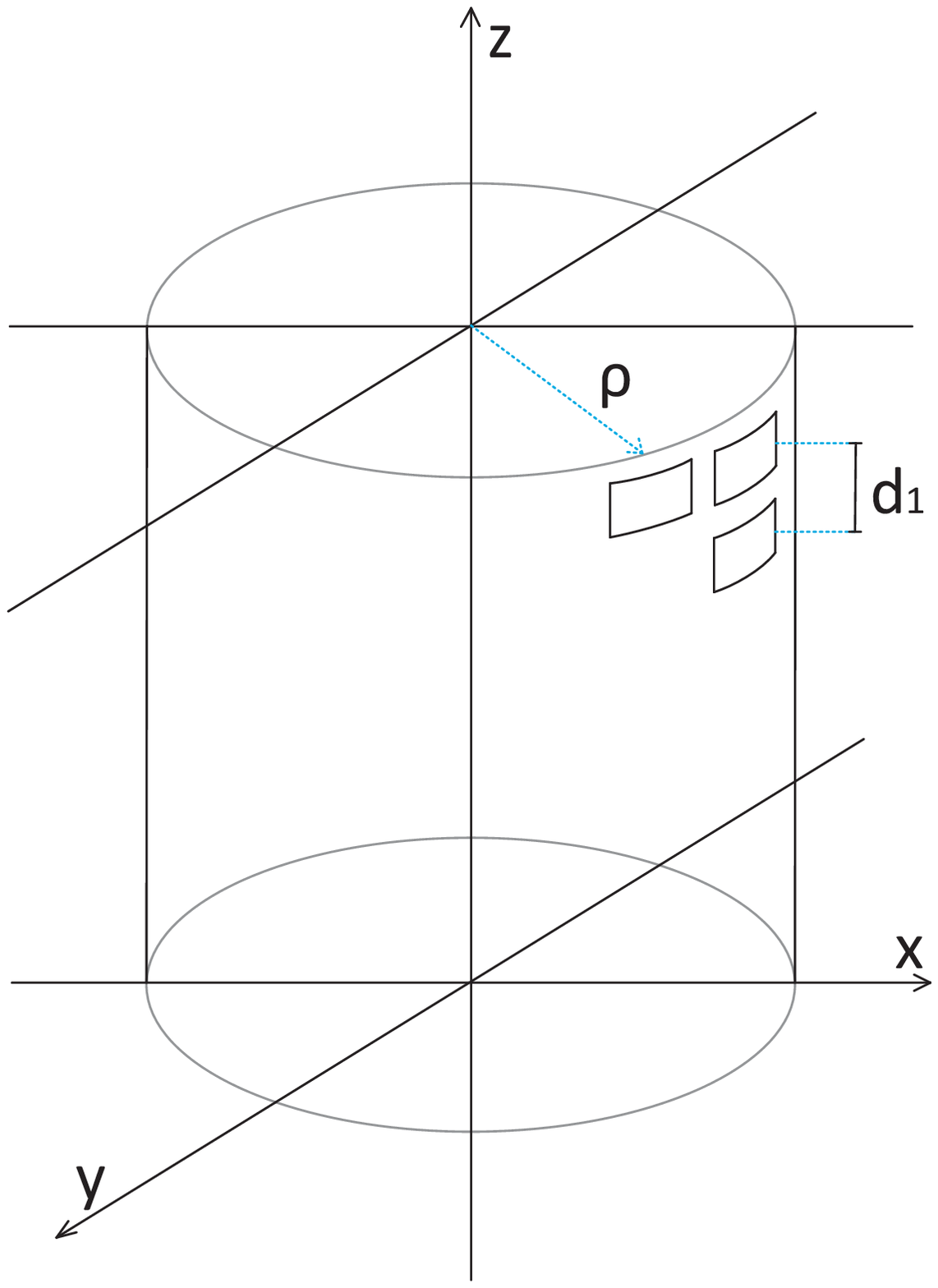}
		\caption{}
                \label{fig:cylindrical}
        \end{subfigure}
	\caption{(a) URA. (b) Cylindrical Array.}
\end{figure}

When large numbers of antennas are present, antenna topologies should exploit $x,y$ and $z$ dimensions. In Section \ref{sec:Methodology}, we derive spatial correlation approximations for the two antenna array topologies: URA and cylindrical, depicted in Figures \ref{fig:ura} and \ref{fig:cylindrical}, respectively. In Figure \ref{fig:ura}, the URA is geometrically positioned on the $y,z$ plane, where adjacent antennas are separated by $d_{1}$ wavelengths on the $z$ axis and $d_{2}$ wavelengths on the $y$ axis. Adjacent antennas of the cylindrical array, in Figure \ref{fig:cylindrical}, are separated by a displacement of $d_{1}$ wavelengths on the $z$ axis and located at a radius of $\rho $ wavelengths from cylinder center, with respect to the $x,y$ plane. 
\par
For both antenna topologies, we express the spatial correlation as the Kronecker product of the azimuth and zenith domain correlations \cite{YING}
\begin{equation}
	\bR=\bR_{\phi }\otimes \bR_{\theta } \label{ying},
\end{equation}
and derive the corresponding correlation coefficients independently in Section \ref{sec:Methodology}. Thus, we consider the URA correlation matrix as the Kronecker product of an $A$-element ULA in the $z$ dimension (zenith domain) and a $B$-element ULA on the $x,y$ plane (azimuth domain), where $M=A\times B$. Likewise, we consider the cylindrical array correlation as the Kronecker product of an $A$-element ULA in the $z$ dimension and a $B$-element UCA on the $x,y$ plane.
\par
%
%
\par
The correlation coefficient between the $a$th and $a'$th antenna in the zenith domain can then by given by \cite{FORENZA}
\begin{equation}
	\bR_{\theta (a,a')} = \int{\textrm{e}^{j[\Theta _{a}(\Delta \theta )-\Theta _{a'}(\Delta \theta )]}p_{\Delta \theta }(\Delta \theta )d\Delta \theta } \label{R_theta_def},
\end{equation}
where $\Theta _{a}(\Delta \theta )$ is the zenith domain phase shift of the $a$th antennas AOD with respect to a reference antenna, $p_{\Delta \theta }(\Delta \theta )$ is the zenith domain AOD offset PDF relative to the mean zenith AOD of the cluster. Likewise, the correlation coefficient between the $b$th and $b'$th antenna element in the azimuth domain is
\begin{align}
	\bR_{\phi (b,b')} &= \int{\int{\textrm{e}^{j[\Phi _{b}(\Delta \phi ,\Delta \theta )-\Phi _{b'}(\Delta \phi ,\Delta \theta )]}}} \notag \\
	&~~~~~~~~~~~~~~~~~~~\times p_{\Delta \phi }(\Delta \phi )p_{\Delta \theta }(\Delta \theta )d\Delta \phi d\Delta \theta \label{R_phi_def},
\end{align}
where $\Phi _{b}(\Delta \phi ,\Delta \theta )$ is the azimuth domain phase shift of the $b$th antennas AOD with respect to a reference antenna, $p_{\Delta \phi }(\Delta \phi )$ is the azimuth AOD offset PDF relative to the mean azimuth angle of the cluster. Note, we assume the azimuth and zenith AOD offset PDF are independent, i.e., $p_{\Delta \phi ,\Delta \theta }(\Delta \phi ,\Delta \theta )=p_{\Delta \phi }(\Delta \phi )p_{\Delta \theta }(\Delta \theta )$. 
\par
We model the azimuth and zenith offset AOD PDFs from measured values described by 3GPP \cite{3GPP}. Thus, the azimuth AOD offset, relative to the mean AOD, is modeled as a Wrapped Gaussian PDF, given by \cite{3GPP}
\begin{equation}
	p_{\Delta \phi }(\Delta \phi ) = \left\{ \begin{array}{l l}
	\frac{1}{\sigma _{\Delta \phi }\sqrt{2\pi }}\sum\limits_{i=-\infty }^{\infty }{\textrm{e}^{-\frac{(\Delta \phi +2\pi i)^{2}}{2\sigma _{\Delta \phi }^{2}}}}  & \Delta \phi \in [-\pi ,\pi ) \\
	0 & \textrm{otherwise}
	\end{array}\right. \label{phi_pdf},
\end{equation}
where $\sigma _{\Delta \phi }$ is the standard deviation (SD) of $\Delta \phi $. Similarly, the zenith AOD offset, relative to the mean AOD, is modeled by a Laplacian PDF, given by \cite{3GPP}
\begin{equation}
	p_{\Delta \theta }(\Delta \theta ) = \left\{ \begin{array}{l l}
	\frac{\kappa }{\sqrt{2}\sigma _{\Delta \theta }}\textrm{e}^{-\left| \frac{\sqrt{2}\Delta \theta }{\sigma _{\Delta \theta }}\right| }   & \Delta \theta \in [-\pi ,\pi ) \\
	0 & \textrm{otherwise}
	\end{array}\right. \label{theta_pdf},
\end{equation}
where $\sigma _{\Delta \theta }$ is the SD of $\Delta \theta $ and $\kappa = 1/(1-\textrm{e}^{-\sqrt{2}\pi /\sigma _{\Delta \theta }})$ normalizes the PDF.
\subsection{Convergence Metrics}
\label{subsec:ConvMetrics}
In order to study the effects of antenna topology on the convergence of massive MIMO properties, we consider the convergence metrics used in \cite{SMITH}. We evaluate the convergence of $\bW = \frac{1}{M}\bH^{\textrm{T}}\bH^{\ast }$ by examining a number of well known properties of $\bW$ and a deviation matrix $\bE=\bW-\bI_{K}$, where $\bI_{K}$ is the $K\times K$ identity matrix. Letting $\lambda _{1},\lambda _{2},\ldots ,\lambda _{K}$ denote the eigenvalues of $\bW$, we consider: $\lambda $ range, Mean Absolute Deviation (MAD) and Diagonal Dominance, defined respectively as
\begin{align}
	\lambda \textrm{ range} &= \lambda _{\textrm{max}}(\bW)-\lambda_ {\textrm{min}}(\bW), \label{lambdarange} \\
	\textrm{MAD}(\mathbf{E}) &= \frac{1}{K^{2}}\sum_{i=1,j=1}^{K}|\mathbf{E} _{ij}| , \label{mad} \\
	\textrm{Diagonal Dominance} &= \frac{\sum_{i=1}^{K}{\bW_{ii}}}{ \sum_{i=1}^{K}{\sum_{j=1,j\neq i}^{K}{|\bW_{ij}|}}}. \label{diagonaldominance}
\end{align}
These metrics will be evaluated via simulation for a number of system scenarios in Section \ref{sec:Numerical_Results}.
%
\section{Methodology}
\label{sec:Methodology}
We now derive approximations for the spatial correlation of URA and cylindrical array topologies. In each case, for simplicity we assume no antenna mechanical downtilt. As per \eqref{ying}, we express the spatial correlation as the Kronecker product of the azimuth and zenith domain correlations, and derive the corresponding correlation coefficients independently. 
\subsection{Uniform Rectangular Array (URA)}
\label{subsec:URA}
We first consider the correlation in the zenith domain, $\bR_{\theta }$, of the URA, depicted in Figure \ref{fig:ura} and geometrically described in Section \ref{subsec:Channel_Model}. The zenith domain phase shift of the departing wavefront from the $a$th antenna element, relative to a reference antenna, can be expressed as \cite{BALANIS}
\begin{equation}
	\Theta _{a}(\Delta \theta )=kd_{1}a\cos (\theta +\Delta \theta) \label{phase_shift},
\end{equation}
where $k$ is the wavenumber. The angle in \eqref{phase_shift} can be expanded with a first-order Taylor series, while assuming $\Delta \theta \approx 0$, to give
\begin{equation}
	\cos (\theta +\Delta \theta) \approx \cos (\theta )-\Delta \theta \sin (\theta ).
\end{equation}
Since $p_{\Delta \theta }(\Delta \theta )$ is zero outside the range $[-\pi ,\pi )$, the PDFs integration can be taken over $[-\infty ,+\infty ]$. From \eqref{R_theta_def} we then have
\begin{align}
	&\bR_{\theta (a,a')} \approx \int_{-\infty }^{\infty }{\textrm{e}^{jkd_{1}(a-a')[\cos (\theta )-\Delta \theta \sin (\theta )]}p_{\Delta \theta }(\Delta \theta )d\Delta \theta } \\
	&= \textrm{e}^{jkd_{1}(a-a')\cos (\theta )} \int_{-\infty }^{\infty }{\textrm{e}^{-jkd_{1}(a-a')\sin (\theta )\Delta \theta }} \notag 
\end{align}
\begin{align}
	&\times \frac{\kappa }{\sqrt{2}\sigma _{\Delta \theta }}\textrm{e}^{-\left| \frac{\sqrt{2}\Delta \theta }{\sigma _{\Delta \theta }}\right| }d\Delta \theta   \\
	&= \textrm{e}^{jkd_{1}(a-a')\cos (\theta )}\mathcal{F}_{\omega }\left\{ \frac{\kappa }{\sqrt{2}\sigma _{\Delta \theta }}\textrm{e}^{-\left| \frac{\sqrt{2}\Delta \theta }{\sigma _{\Delta \theta }}\right| }\right\} \label{URA_el_2},
\end{align}
where $\mathcal{F}_{\omega }$ denotes the Fourier transform evaluated at $\omega =kd_{1}(a-a')\sin (\theta )$. Solving the Fourier transform in \eqref{URA_el_2}, we have the approximation of the correlation coefficient for the zenith domain of a URA as
\begin{align}
	&\bR_{\theta (a,a')} \approx \frac{\kappa \textrm{e}^{jkd_{1}(a-a')\cos (\theta )}}{1+\frac{\sigma _{\Delta \theta }^{2}}{2}[kd_{1}(a-a')\sin (\theta )] ^{2}} \label{R_theta}.
\end{align}
\par
We now consider the correlation in the azimuth domain. The azimuth domain phase shift of the departing wavefront from the $b$th antenna element, relative to a reference antenna, can be expressed as \cite{BALANIS}
\begin{equation}
	\Phi _{b}(\Delta \phi ,\Delta \theta ) = kd_{2}b\cos (\phi +\Delta \phi )\sin (\theta +\Delta \theta ) \label{URA_az_6}.
\end{equation}
We can then use a first-order Taylor series expansion, while assuming $\Delta \phi \approx 0$, to express \eqref{URA_az_6} as
\begin{equation}
	\cos (\phi +\Delta \phi )\sin (\theta +\Delta \theta ) \approx \sin (\theta +\Delta \theta )\left[ \cos (\phi )-\Delta \phi \sin (\phi )\right] \label{URA_az_2}.
\end{equation}
Since both $p_{\Delta \phi }(\Delta \phi )$ and $p_{\Delta \theta }(\Delta \theta )$ are zero outside the range $[-\pi ,\pi )$, the integration can be taken over $[-\infty ,+\infty ]$. From \eqref{R_phi_def}, the correlation coefficient between two antenna elements $b$ and $b'$ is then given by
\begin{align}
	&\bR_{\phi (b,b')} \approx \int_{\infty }^{\infty }{\textrm{e}^{jkd_{2}(b-b')\sin (\theta +\Delta \theta )\cos (\phi )}} \notag
\end{align}
\begin{align}
	&\times \left[ \int_{-\infty }^{\infty }{\textrm{e}^{-jkd_{2}(b-b')\sin (\theta +\Delta \theta )\sin (\phi )\Delta \phi }p_{\Delta \phi }(\Delta \phi )d\Delta \phi }\right] \notag \\
	&~~~~~~~~~~~~~~~~~~~~~~~~~~~~~~~~~~~~~~~~~~~~\times p_{\Delta \theta }(\Delta \theta )d\Delta \theta  \label{URA_az_3}.
\end{align}
Considering the integral with respect to $\Delta \phi $, embedded in \eqref{URA_az_3}, we have
\begin{align}
	&\int_{-\infty }^{\infty }{\textrm{e}^{-jkd_{2}(b-b')\sin (\theta +\Delta \theta )\sin (\phi )\Delta \phi }p_{\Delta \phi }(\Delta \phi )d\Delta \phi } \notag \\
	&= \int_{-\infty }^{\infty }{\textrm{e}^{-jkd_{2}(b-b')\sin (\theta +\Delta \theta )\sin (\phi )\Delta \phi }} \notag \\
	&~~~~~~~~~~~~~~~~~~~~~~~\times \sum_{i=-\infty }^{\infty }{\frac{1}{\sigma _{\Delta \phi }\sqrt{2\pi }}\textrm{e}^{-\frac{(\Delta \phi +2\pi i)^{2}}{2\sigma _{\Delta \phi }^{2}}}}d\Delta \phi  \\
	&= \mathcal{F}_{\omega }\left\{ \sum_{i=-\infty }^{\infty }{\frac{1}{\sigma _{\Delta \phi }\sqrt{2\pi }}\textrm{e}^{-\frac{(\Delta \phi +2\pi i)^{2}}{2\sigma _{\Delta \phi }^{2}}}}\right\} \label{URA_el_7},
\end{align}
where the Fourier transform in \eqref{URA_el_7} is evaluated at $\omega =kd_{2}(b-b')\sin (\theta +\Delta \theta )\sin (\phi )$. Evaluating the Fourier transform, one obtains
\begin{align}
	&\bR_{\phi (b,b')}\approx \textrm{e}^{jkd_{2}(b-b')\sin (\theta )\cos (\phi )}\frac{\kappa }{\sqrt{2}\sigma _{\Delta \theta }} \notag \\
	&\times \sum_{i=-\infty }^{\infty }{\textrm{e}^{\frac{-1}{2(1-j2\pi i)}\left[ \sigma _{\Delta \phi }kd_{2}(b-b')\sin (\theta )\sin (\phi )\right] ^{2}}} \notag 
\end{align}
\begin{align}
	&~~~~~\times \mathcal{F}_{\omega }\left\{ \textrm{e}^{\frac{-1}{2(1-j2\pi i)}\left[ \sigma _{\Delta \phi }kd_{2}(b-b')\cos (\theta )\sin (\phi )\right] ^{2}(\Delta \theta )^{2}}\right\} ,\label{URA_az_5}
\end{align}
where steps from \eqref{URA_el_7} to \eqref{URA_az_5} are given in the Appendix and the Fourier transform in \eqref{URA_az_5} is evaluated at 
\begin{equation}
	\omega = \left\{ \begin{array}{rcl}
	C_{i}-\frac{\sqrt{2}}{\sigma _{\Delta \theta }} & \mbox{for} & \Delta \theta \geq 0 \\
	C_{i}+\frac{\sqrt{2}}{\sigma _{\Delta \theta }} & \mbox{for} & \Delta \theta < 0
	\end{array}\right. \label{omega},
\end{equation}
where 
\begin{align}
	&C_{i}=jkd_{2}(b-b')\cos (\theta ) \notag \\
	&\times \left[ \cos (\phi )-\frac{1}{j+2\pi i}\sigma _{\Delta \phi }^{2}kd_{2}(b-b')\sin ^{2}(\phi )\sin (\theta )\right] .
\end{align}
Evaluating the Fourier transform in \eqref{URA_az_5}, we have the approximation of the correlation coefficient for the azimuth domain of a URA as
\begin{align}
	&\bR_{\Delta \phi (b,b')} \approx \frac{\kappa }{\sqrt{2}\sigma _{\Delta \theta }}\textrm{e}^{jkd_{2}(b-b')\sin (\theta )\cos (\phi )} \notag \\
	&\times \sum_{i=-\infty }^{\infty }{\textrm{e}^{\frac{-1}{2(1-j2\pi i)}\left[ \sigma _{\Delta \phi }kd_{2}(b-b')\sin (\theta )\sin (\phi )\right] ^{2}}} \notag \\
	&~~~~~~~~~~~~~~~~~~~~~~~~\times \textrm{e}^{\frac{-\omega ^{2}}{\frac{2}{1-j2\pi i}\left[ \sigma _{\Delta \phi }kd_{2}(b-b')\cos (\theta )\sin (\phi )\right] ^{2}}}, \label{R_phi}
\end{align}
with $\omega $ given in \eqref{omega}. The spatial correlation approximation for the URA is then as per \eqref{ying}, with $\bR_{\theta }$ and $\bR_{\phi }$ given in \eqref{R_theta} and \eqref{R_phi} respectively. 
\subsection{Cylindrical Array}
\label{subsec:cylindrical_array}
Similar to the analysis for URA, the spatial correlation of a cylindrical array, shown in Figure \ref{fig:cylindrical}, can be broken down into the Kronecker product of the azimuth and zenith domains, described in Section \ref{subsec:Channel_Model}. The zenith domain phase shift of the $a$th antenna element, $\Theta _{a}(\Delta \theta )$, and the correlation coefficient between antennas $a$ and $a'$, $\bR_{\theta }(a,a')$, are identical to the URA case and are thus given in \eqref{phase_shift} and \eqref{R_theta} respectively. In the azimuth domain, the phase shift of the $b$th antenna element, relative to a reference antenna is given by \cite{BALANIS}
\begin{equation}
	\Phi _{b}(\Delta \phi ,\Delta \theta ) = k\rho b\cos ((\phi -\phi _{b})+\Delta \phi )\sin (\theta +\Delta \theta ) \label{CA_az_1},
\end{equation}
where $\phi -\phi _{b}$ is the angle between the incident ray projected onto the $x,y$ plane and the $b$th antenna, relative to the the circle center. Note that the azimuth domain phase shift in \eqref{CA_az_1} is analogous to \eqref{URA_az_6}, with $\rho =d_{2}$ and $\phi -\phi _{b}=\phi $. Therefore, the spatial correlation approximation for the cylindrical array is given in \eqref{ying}, where 
\begin{align}
	&\bR_{\phi (b,b')} \approx \frac{\kappa }{\sqrt{2}\sigma _{\Delta \theta }}\textrm{e}^{jk\rho (b-b')\sin (\theta )\cos (\phi -\phi _{b})} \notag \\
	&\times \sum_{i=-\infty }^{\infty }{\textrm{e}^{\frac{-1}{2(1-j2\pi i)}\left[ \sigma _{\Delta \phi }k\rho (b-b')\sin (\theta )\sin (\phi -\phi _{b})\right] ^{2}}} \notag \\
	&~~~~~~~~~~~~~~~~~~~~~\times \textrm{e}^{\frac{-\omega ^{2}}{\frac{2}{1-j2\pi i}\left[ \sigma _{\Delta \phi }k\rho (b-b')\cos (\theta )\sin (\phi -\phi _{b})\right] ^{2}}},
\end{align}
where $\omega $ is given in \eqref{omega}, with 
\begin{align}
	&C_{i}=jk\rho (b-b')\cos (\theta ) ~\times \notag 
\end{align}
\begin{align}
	&\left[ \cos (\phi -\phi _{b})-\frac{1}{j+2\pi i}\sigma _{\Delta \phi }^{2}k\rho (b-b')\sin ^{2}(\phi -\phi _{b})\sin (\theta )\right] ,
\end{align}
and $\bR_{\theta }$ is given in \eqref{R_theta}.
%
\section{Numerical Results}
\label{sec:Numerical_Results}
We consider the convergence metrics, described in Section \ref{subsec:ConvMetrics}, in order to determine how many antennas are required for observable massive MIMO properties in a spatially correlated environment. System parameters are presented in Table \ref{system_parameters}.
\begin{table}[h]
\centering
\begin{tabular}{cc}
\hline
\multicolumn{1}{c}{\textbf{Parameter}} & \textbf{Value} \\ \hline
	Frequency (GHz) & 2.6 \\ \hline
	X-pol parameter, $\sqrt{\delta }$ & $0.1$ \\ \hline
	Azimuth AOD offset PDF, $p_{\Delta \phi }(\Delta \phi )$ & Wrapped Gaussian \\ \hline
	Zenith AOD offset PDF, $p_{\Delta \theta }(\Delta \theta )$ & Laplacian \\ \hline
	AOD cluster mean, \{$\phi, \theta $\} ($\textrm{log}_{10}([^{\circ}])$) & 0.7 \\ \hline
	AOD offset SD, \{$\sigma _{\Delta \phi },\sigma _{\Delta \theta }$\} ($\textrm{log}_{10}([^{\circ}])$) & -0.3 \\ \hline
\end{tabular}
\caption{System Parameters}
\label{system_parameters}
\end{table}
\subsection{Convergence Properties}
\label{subsec:Convergence_Properties}
%
\begin{figure}[ht]
\centering\includegraphics[width=1\columnwidth]{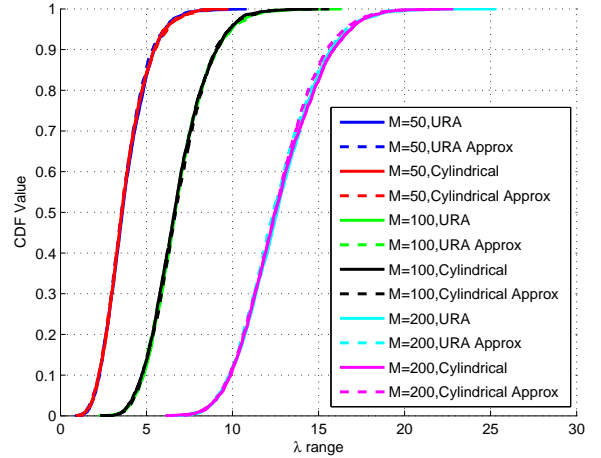}
\raisecaption\caption{$\lambda $ range CDF for $\frac{M}{K}=\alpha =10$}
\label{lambda_range}
\end{figure}
In Figure \ref{lambda_range}, we show the cumulative distribution function (CDF) of the $\lambda $ range, given in \eqref{lambdarange}, for varying $M$. The eigenvalues of $\bW$ are generated using $\bH$, given in \eqref{channel_matrix}, where the simulated CDFs are generated from instantaneous array factors, while the approximations are generated from \eqref{ying}. It is seen that as we increase the number of transmit antennas, $M$, the median value of the $\lambda $ range CDF increases, rather than converging to an equal eigenvalued channel, which is observed in \cite{SMITH} and shown in \cite{RUSEK} for an i.i.d. channel. This is due to an increase in the dominant eigenvalue of $\bW$, resulting from such a narrow angle spread. From Figure \ref{lambda_range}, we see that for all values of $M$, our derived expressions approximate the spatial correlation well. Thus, we use the approximations for all results following.
%
\begin{figure}[ht]
\centering\includegraphics[width=1\columnwidth]{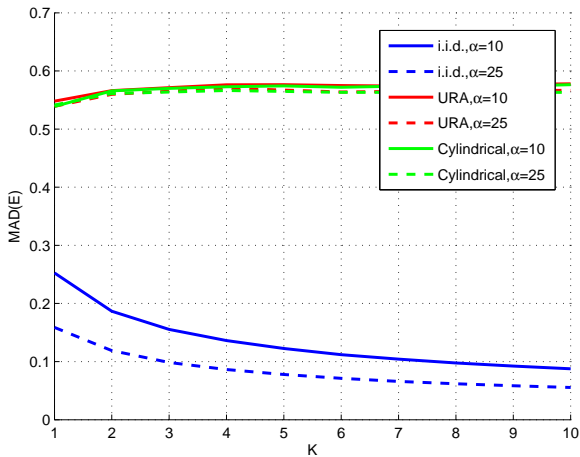}
\raisecaption\caption{MAD($\bE$) vs $K$, with $\frac{M}{K}=\alpha =10$ and $25$}
\label{MAD}
\end{figure}
\par
In Figure \ref{MAD} we plot the average of MAD($\bE$), given in \eqref{mad}, versus $K$ for $\alpha =10$ and $25$. We observe very different behaviour between the correlated and i.i.d. scenarios. In the i.i.d. case, MAD($\bE$) converges slowly to zero for increasing $K$, with a quicker convergence rate for larger $\alpha $. Note that the average MAD($\bE$) value is plotted in Figure \ref{MAD} and the only way for this to converge to zero is for all $\bW$ matrices to be close to $\bI_{K}$. Hence, only in the i.i.d. case does each $\bW$ become close to $\bI_{K}$ as $K$ increases. 
%
\begin{figure}[ht]
\centering\includegraphics[width=1\columnwidth]{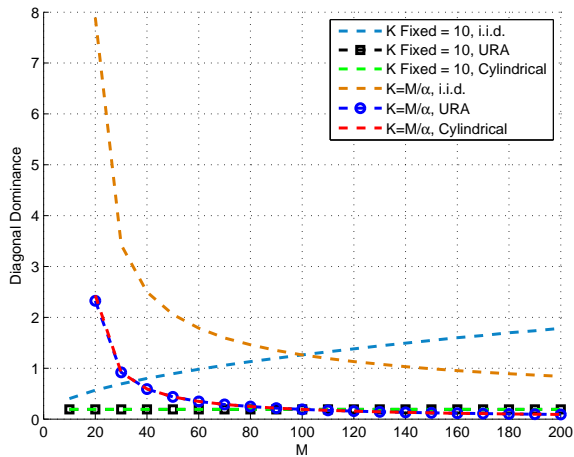}
\raisecaption\caption{Diagonal Dominance vs $M$, with $K$ fixed and $K=\frac{M}{\alpha }$}
\label{diagonal_dominance}
\end{figure}
\par
Figure \ref{diagonal_dominance} shows the diagonal dominance of $\bW$ (size $K\times K$) given in \eqref{diagonaldominance}, as a function of the number of transmit antennas, $M$. We see that the correlated scenario has \emph{similar} trends as the i.i.d. case. As with the results in \cite{SMITH} for $K=\frac{M}{\alpha }$, the diagonal dominance  decays. This follows as the number of off-diagonals, which grow at a rate $\approx K^{2}$, increase the denominator of \eqref{diagonaldominance} at a faster rate than the numerator. In the correlated scenario, the diagonal dominance approaches zero much faster than the i.i.d. case. Here, while the diagonal elements converge to a mean of 1, the large number of off-diagonals converge to a \emph{non-zero} mean, dominating the diagonal elements. On the other hand, for fixed $K$, the sum of the diagonal elements increase by a greater proportion than the off-diagonal elements as $M$ is increased, resulting in a more diagonally dominant i.i.d. $\bW $. In the correlated case the diagonal dominance converges quickly to small non-zero value.
\par
Considering Figures \ref{lambda_range}, \ref{MAD} and \ref{diagonal_dominance}, we conclude that while for i.i.d. channels, the massive MIMO metrics converge for a very large numbers of antennas, desirable massive MIMO properties are degraded with spatial correlation present (with small angle spreads).
\subsection{Convergence Properties of MF Precoder}
\label{subsec:Convergence_Properties_MF}
We now explore the impact of massive MIMO antenna topologies on MF SINR performance and convergence to limiting values. The MF SINR of the $i$th user is given by
\begin{equation}
	\textrm{SINR}_{i} = \frac{\frac{\rho _{\textrm{d}}}{K\gamma }|\bh_{i}^{\textrm{T}}\bh_{i}^{\ast }|^{2}}{1 + \frac{\rho _{\textrm{d}}}{K\gamma }\sum_{j=1,j\neq i}^{K}{\bh_{i}^{\textrm{T}}\bh_{j}^{\ast }\bh_{j}^{\textrm{T}}\bh_{i}^{\ast }}} \label{sinr},
\end{equation}
where $\rho _{\textrm{d}}$ is the DL transmit signal to noise ratio (SNR), $\bh_{i}$ denotes the $i$th column of $\bH$, and $\gamma = \textrm{tr}(\bH^{\textrm{T}}\bH^{\ast })/K$ is the power normalization factor.
%
\begin{figure}[ht]
\centering\includegraphics[width=1\columnwidth]{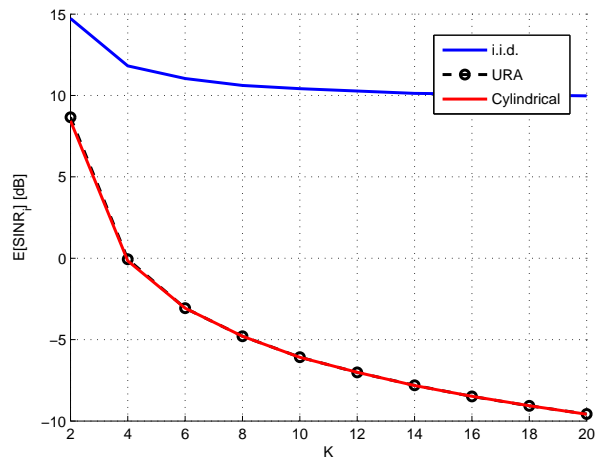}
\raisecaption\caption{MF SINR for, $\frac{M}{K}=\alpha =10$, $\rho _{\textrm{d}}=10$ dB}
\label{mf_sinr}
\end{figure}
\par
In Figure \ref{mf_sinr} we plot the expected value of \eqref{sinr}. It can be observed that there is a huge reduction in $\mathbb{E}[\textrm{SINR}_{i}]$ performance by introducing correlation of which antenna topology has almost no effect on. The MF SINR for an i.i.d. scenario rapidly converges to its limiting value, whereas in a correlated case, $\mathbb{E}[\textrm{SINR}_{i}]$ converges to limiting values for $M>200$. We conclude that linear precoders are highly sub-optimal for massive MIMO systems in spatially correlated environments with small angle spreads.
%
\section{Conclusion}
\label{sec:Conclusion}
In this paper we develop approximations for the spatial correlation of URA and cylindrical antenna arrays, from a 3D channel model. Using derived approximations, we show that large spatial correlation, due to small angle spread, destroys the convergence of massive MIMO properties with increasing number of antennas. Furthermore, the impact of massive MIMO antenna topology considered is shown to be negligible. MF SINR performance and convergence rate is explored for the two antenna topologies, showing the detrimental effect of correlation, with respect to an i.i.d. channel. It should be noted that if we increased inter-element spacings of the antenna arrays relative to the angle spread, we would see a decrease in correlation and the convergence metrics would approach the i.i.d. case. We leave this to future work.
\section{Future Work}
\label{sec:Future_Work}
In the future, we aim at deriving correlation matrices for various antenna topologies which do not rely on small-angle distributions. Also, we aim to incorporate the effects of mutual coupling, due to the proximity of antennas as electrical components, important for the analysis of large antenna arrays. Furthermore, an investigation into how the various antenna spacing parameters influence the correlation for the antenna topologies will give insight into the structure of the correlation matrices and how to carefully design an antenna array.
%
%
\appendix
\label{app:A}
From \eqref{URA_el_7}, we have 
\begin{align}
	&\bR_{\phi (b,b')}\approx \int_{\infty }^{\infty }{\textrm{e}^{jkd_{2}(b-b')\sin (\theta +\Delta \theta )\cos (\phi )}} \sum_{i=-\infty }^{\infty }{\textrm{e}^{\frac{-1}{2(1-j2\pi i)}}} \notag \\
	&~~~~~~~~~~~^{\times \left[ \sigma _{\Delta \phi }kd_{2}(b-b')\sin (\theta +\Delta \theta )\sin (\phi )\right] ^{2}} p_{\Delta \theta }(\Delta \theta )d\Delta \theta  \label{tmp_1} \\
	&\approx \int_{\infty }^{\infty }{\textrm{e}^{jkd_{2}(b-b')\left[ \sin (\theta )+\Delta \theta \cos (\theta )\right] \cos (\phi )}} \sum_{i=-\infty }^{\infty }{\textrm{e}^{\frac{-1}{2(1-j2\pi i)}}} \notag \\
	&~~~^{\times \left[ \sigma _{\Delta \phi }kd_{2}(b-b')\left[ \sin (\theta )+\Delta \theta \cos (\theta )\right] \sin (\phi )\right] ^{2}} p_{\Delta \theta }(\Delta \theta )d\Delta \theta \label{URA_az_4} \notag \\
	&=\textrm{e}^{jkd_{2}(b-b')\sin (\theta )\cos (\phi )}\sum_{i=-\infty }^{\infty }{\textrm{e}^{\frac{-1}{2(1-j2\pi i)}}} \notag \\
	&^{\times \left[ \sigma _{\Delta \phi }kd_{2}(b-b')\sin (\theta )\sin (\phi )\right] ^{2}}\int_{-\infty }^{\infty }{\textrm{e}^{jkd_{2}(b-b')\cos (\theta )}} \notag \\
	&^{\times \left[ \cos (\phi )-\frac{1}{j+2\pi i}\sigma _{\Delta \phi }^{2}kd_{2}(b-b')\sin ^{2}(\phi )\sin (\theta )\right] \Delta \theta } \times \notag \\
	&\textrm{e}^{\frac{-1}{2(1-j2\pi i)}\left[ \sigma _{\Delta \phi }kd_{2}(b-b')\cos (\theta )\sin (\phi )\right] ^{2}(\Delta \theta )^{2}}p_{\Delta \theta }(\Delta \theta )d\Delta \theta \\
	&=\textrm{e}^{jkd_{2}(b-b')\sin (\theta )\cos (\phi )}\frac{\kappa }{\sqrt{2}\sigma _{\Delta \theta }}\sum_{i=-\infty }^{\infty }{\textrm{e}^{\frac{-1}{2(1-j2\pi i)}}} \notag \\
	&^{\times \left[ \sigma _{\Delta \phi }kd_{2}(b-b')\sin (\theta )\sin (\phi )\right] ^{2}}\int_{-\infty }^{\infty }{\textrm{e}^{jkd_{2}(b-b')\cos (\theta )}} \notag \\
	&^{\times \left[ \cos (\phi )-\frac{1}{j+2\pi i}\sigma _{\Delta \phi }^{2}kd_{2}(b-b')\sin ^{2}(\phi )\sin (\theta )\right] \Delta \theta } \times \textrm{e}^{-\frac{\sqrt{2}}{\sigma _{\Delta \theta }}|\Delta \theta |} \notag \\
	&~~~~~~~~\times \textrm{e}^{\frac{-1}{2(1-j2\pi i)}\left[ \sigma _{\Delta \phi }kd_{2}(b-b')\cos (\theta )\sin (\phi )\right] ^{2}(\Delta \theta )^{2}}d\Delta \theta ,
\end{align}
where a Taylor series expansion is used to obtain \eqref{URA_az_4}.
\bibliographystyle{IEEEtran}
\bibliography{bibliography}
\end{document}